\documentclass[useAMS,usenatbib,usegraphicx]{mn2e}
\usepackage{amsmath}
\title[Systematic detection of magnetic fields in massive, late-type supergiants]{Systematic detection of magnetic fields in massive, late-type supergiants\thanks{Based on observations obtained at the Canada-France-Hawaii Telescope (CFHT) which is operated by the National Research Council of Canada, the Institut National des Sciences de l'Univers of the Centre National de la Recherche Scientifique of France,  and the University of Hawaii.}}
\author[Grunhut et al.]{J.H. Grunhut$^{1,2}$\thanks{E-mail: Jason.Grunhut@rmc.ca}, G.A. Wade$^2$, D.A. Hanes$^{1}$, E. Alecian$^3$\\
$^1$Department of Physics, Engineering Physics \& Astronomy, Queen’s University, Kingston, Ontario, Canada, K7L 3N6\\
$^2$Department of Physics, Royal Military College of Canada, P.O. Box 17000, Station Forces, Kingston, Ontario, Canada, K7K 7B4\\
$^3$LAOG, Laboratoire d'Astrophysique de Grenoble, Universit\'{e} Joseph Fourier, Grenoble Cedex, France
}

\begin{document}
\date{\today}
\pagerange{\pageref{firstpage}--\pageref{lastpage}} \pubyear{2009}
\maketitle
\label{firstpage}
\begin{abstract}
We report the systematic detection of magnetic fields in massive ($M>5$\,$M_\odot$) late-type supergiants, using spectropolarimetric observations obtained with ESPaDOnS at the Canada-France-Hawaii Telescope. Our observations reveal detectable Stokes $V$ Zeeman signatures in Least-Squares Deconvolved mean line profiles in one-third of the observed sample of more than 30 stars. The signatures are sometimes complex, revealing multiple reversals across the line. The corresponding longitudinal magnetic field is seldom detected, although our longitudinal field error bars are typically $0.3$~G ($1\sigma$). These characteristics suggest topologically complex magnetic fields, presumably generated by dynamo action. The Stokes~$V$ signatures of some targets show clear time variability, indicating either rotational modulation or intrinsic evolution of the magnetic field. We also observe a weak correlation between the unsigned longitudinal magnetic field and the Ca~{\sc ii} K core emission equivalent width of the active G2Iab supergiant $\beta$~Dra and the G8Ib supergiant $\epsilon$~Gem. 
\end{abstract}


\section{Introduction}
Supergiants are the descendants of massive O and B-type main sequence stars. When a massive star completes its main sequence evolution, it evolves rapidly across the Hertzsprung-Russell (HR) diagram becoming a cool supergiant, characterized by a helium-burning core, and a deep convective hydrogen-burning envelope. This is in strong contrast to their OB main sequence phase, where they are characterized by a convective hydrogen-burning core and a radiative envelope \citep[e.g][]{brown06}. Subsequent evolution of the most massive stars (initial mass $>25$\,$M_{\odot}$) involves the ejection of their envelope in the Wolf-Rayet phase and a dramatic explosion in a type Ib supernova, while the less massive stars (initial mass between about 8 and 25\,$M_{\odot}$) evolve into red supergiants that explode as type II supernovae. Ultimately, the end product is a neutron star or black hole \citep[e.g.][]{crow95, eld08}.

Due to their extended radii, low atmospheric densities, slow rotation and long convective turnover times, supergiants provide an opportunity to study stellar magnetism and activity at the extremes of parameter space. Observations of late-type supergiants (spectral type F and later) report an array of diverse and puzzling activity phenomena. Active late-type supergiants are luminous in X-rays and show emission in chromospheric ultraviolet (UV) Si~{\sc iv} lines, with some stars showing evidence of flaring - phenomena associated with the presence of a corona and magnetic reconnection, presumably resulting from a dynamo-driven magnetic field \citep[e.g.][]{tar02, ayres05a}. However, unlike main sequence dwarfs, supergiants show little, if any, rotation-activity relation and are X-ray deficient by an order of magnitude or more compared to main sequence dwarf stars of similar spectral type \citep[e.g.][]{ayres05b}. There also exists a class of ``non-coronal" or inactive supergiants that exhibit weak or no X-ray emission, UV spectra containing mostly low-temperature chromospheric lines (O~{\sc i}, Mg~{\sc ii}: $T\sim10^4$\,K) and much weaker UV fluxes in C~{\sc iv} chromospheric emission lines \citep[][]{ayres05b}. The position of these non-coronal stars on the HR diagram coincides with the regions where cool (T$\le 2\times10^4$\,K) winds are predicted to dominate the circumstellar environment, potentially masking their X-ray emission \citep[e.g.][]{haisch92, ayres05a}. Another possible explanation for their apparent lack of activity is that magnetic loops might be submerged within the thick extended chromosphere \citep{ayres05a}, resulting in internal absorption of coronal X-rays. Curiously, there also exist some ``hybrid" stars that show the strong C~{\sc iv} UV line emission of active supergiants, but that are weak or undetected in X-rays \cite[e.g][]{ayres05a}.

Observations of the M2 supergiant Betelgeuse \citep[$M=10$\,$M_{\odot}$;][]{carp97} have revealed a chromosphere extending beyond 120 R$_{*}$, as well as evidence of irregular surface brightness fluctuations, which simulations suggest are due to giant convective cells \citep{frey02}. Numerical simulations \citep{dorch04} predict that these convective regions are able to excite a dynamo capable of generating a highly structured surface magnetic field with localised features as strong as 500\,G. In fact, \citet{aur10} recently reported the detection of a magnetic field in this star, with a (disc-integrated) longitudinal field component varying from about 0.5-1.5 G over about one month.

Motivated by the outstanding puzzles associated with the activity of late-type supergiants, the near complete lack of direct constraints on their magnetic fields, and recent successes measuring the magnetic fields of red and yellow giants \citep[evolved intermediate-mass stars; e.g.][]{aur08,aur09,konst09}, we have initiated a program to search for direct evidence of magnetic fields in supergiants (evolved high-mass stars). The data presented in this paper represent stars with detected Zeeman signatures observed as part of a large survey of over 30 supergiants ranging in spectral type from early A- to late M. 

\section{Observations}\label{obs_sec}
Circular polarisation (Stokes~$V$) spectra were obtained with the ESPaDOnS spectropolarimeter, mounted on the 3.6m Canada-France-Hawaii Telescope (CFHT), as part of a larger survey investigating the magnetic properties of bright, late-type supergiants. In addition, one spectrum was acquired with the NARVAL spetropolarimeter on the Bernard Lyot Telescope (TBL). 

Both the ESPaDOnS and NARVAL instruments are optical echelle spectropolarimeters, capable of yielding broad bandpass (370 to 1050 nm), high resolution ($R=68\,000$), polarised spectra. A complete polarisation observation consists of four individual sub-exposures between which the half-wave retarders are rotated back and forth between position angles of -90$^{\circ}$ and +90$^{\circ}$ in order to reduce first-order systematic errors in the polarisation analysis. The extraction of the ESPaDOnS and NARVAL spectra, along with wavelength calibration and continuum normalisation, was accomplished using the Upena pipeline running Libre-ESpRIT \citep[][]{don97}, a dedicated automatic reduction package installed at both CFHT and TBL. To avoid saturation, we often obtained multiple consecutive sequences of observations of a target that we co-added after extraction. The median signal-to-noise ratio (S/N) of the co-added observations (computed from photon counting statistics, per 1.8 km\,s$^{-1}$ spectral pixel in the $V$-band) was about 2500, but ranged as high as 4700.

More than 30 stars have been observed as part of a survey of the brightest supergiants. The current sample contains 4 A-type stars, 8 F-type stars, 11 G-type stars, 7 K-type stars, and 3 M-type stars. In this paper we report the current results of this survey, with a focus on the 9 stars in which clear Stokes~$V$ Zeeman signatures have been detected. The journal of observations for the observed stars is provided in Table \ref{obs_tab}. 

\begin{table*}
\centering
\caption{Journal of ESPaDOnS/NARVAL observations. Columns 1-7 list the target name, the HD number, the spectral type (from SIMBAD), the observation date, the exposure time, the peak signal-to-noise ratio (S/N) in the $V$-band in the observed spectrum, and the mean S/N ratio (per 1.8 km~s$^{-1}$ velocity bin) in the LSD Stokes~$V$ profile, for each observation. Columns 8-10 list the mean longitudinal field measurement from the LSD profiles, the false alarm probability (FAP), and the Stokes~$V$ detection diagnosis (DD=Definite Detection, MD=Marginal Detection, ND=No Detection) as described by \citet{don97}. The first 9 rows represent stars in which we observe clear Zeeman signatures in Stokes~$V$, while the next 6 rows represent stars with suspected Zeeman signatures. The last rows represent stars with no visible Zeeman signature in Stokes~$V$.}
\begin{tabular}{lrrlrrrrrr}
\hline
\hline
Name & \multicolumn{1}{c}{HD} & Spec.& \multicolumn{1}{c}{Obs.} & \multicolumn{1}{c}{Exp. Time} &\multicolumn{1}{c}{S/N} & \multicolumn{1}{c}{S/N} & \multicolumn{1}{c}{$\langle B_{\rm z}\rangle \pm \sigma_B$} & \multicolumn{1}{c}{FAP} & \multicolumn{1}{c}{Det.}\\
\ & \ & Type & \multicolumn{1}{c}{Date} & \multicolumn{1}{c}{(s)} & \multicolumn{1}{c}{$V$} & \multicolumn{1}{c}{LSD $V$} & \multicolumn{1}{c}{(G)} & \ & \multicolumn{1}{c}{Flag} \\ 
\hline
$\alpha$ Lep & 36673 & F0Ib & 2009-09-27 & 3$\times$4$\times$32 & 2709 & 65870 & $0.03\pm0.37$ & 2.420E-07 & DD \\
$\alpha$ Per & 20902 & F5Iab & 2010-01-26 & 8$\times$4$\times$15 & 4661 & 130558 & $0.82\pm0.37$ &  $<$1E-16 & DD\\
$\eta$ Aql & 187929 & F6Iab & 2009-09-05 & 3$\times$4$\times$111 & 2488 & 61608 & $-0.23\pm0.75$ & 6.985E-08 & DD\\
$\beta$ Dra & 159181 & G2Iab & 2010-03-05 & 6$\times$4$\times$36 & 4526 & 131246 & $-1.16\pm0.25$ & $<$1E-16 & DD\\
$\xi$ Pup & 63700 & G6Ia & 2010-01-26 & 4$\times$4$\times$61 & 3543 & 112130 & $0.24\pm0.28$ & $<$1E-16 & DD\\
$\epsilon$ Gem & 48329 & G8Ib & 2010-01-25 & 5$\times$4$\times$43 & 4262 & 130843 & $-0.14\pm0.19$ & $<$1E-16 & DD\\
$c$ Pup & 63032 & K2.5Ib-II & 2009-11-28 & 3$\times$4$\times$55 & 2188 & 65948 & $1.10\pm0.39$& 1.410E-06 & DD\\
32 Cyg & 192909 & K3Ib+ & 2009-10-02 & 2$\times$4$\times$103 & 1530 & 50481 & 1.16$\pm$0.49 & 2.053E-04 & MD \\
$\lambda$ Vel & 78647 & K4.5Ib-II & 2009-05-07 & 10$\times$4$\times$5 & 2355 & 75431 & $1.72\pm0.33$ & 5.995E-15 & DD\\
\hline
$\beta$ Aqr & 204867 & G0Ib & 2009-09-06 & 3$\times$4$\times$164 & 2673 & 80881 & $0.33\pm0.31$ & 1.138E-02 & ND\\
$\alpha$~Aqr & 209750 & G2Ib & 2009-09-06 & 3$\times$4$\times$43 & 3313 & 77144 & 0.47$\pm$0.36 &  2.607E-03 & ND \\
$d$~Cen & 117440 & G9Ib & 2010-06-04 & 3$\times$4$\times$73 & 2464 & 76850 & 0.27$\pm$0.26 & 4.313E-01 & ND\\
$\xi$~Cyg & 200905 & K4.5Ib-II & 2009-10-02 & 2$\times$4$\times$80 & 1640 & 54037 & $-0.46\pm0.36$ & 2.850E-02 & ND\\
$\alpha$~Ori & 39801 & M2Iab & 2009-09-28$^1$ & 6$\times$4$\times$1 & 2216 &62459 & -0.80$\pm$0.45 & 3.285E-01 & ND\\
$\sigma$~CMa & 52877 & M1.5Iab & 2009-10-02 & 2$\times$4$\times$49 & 2033 & 66480 & 0.61$\pm$0.32 &  3.060E-01 & ND\\
\hline
$\eta$~Leo & 87737 & A0Ib & 2009-12-07 & 3$\times$4$\times$93 & 3069 & 48084 & 0.03$\pm$1.71 & 9.896E-01 & ND \\
$\alpha$~Cyg & 197345 & A2Iae & 2009-09-05 & 3$\times$4$\times$44 & 2861 & 60421 & 2.47$\pm$2.12 &  4.378E-01 & ND\\
$L$~Pup & 62623 & A3Iab & 2009-12-09 & 4$\times$4$\times$104 & 2749 & 55470 & -9.76$\pm$3.49 & 5.837E-01 & ND\\
$\epsilon$~Aur & 31964 & A8Iab & 2009-09-28 & 3$\times$4$\times$47 & 2671 & 56858 & 1.27$\pm$1.52 & 5.013E-01 & ND \\
$LN$~Hya & 112374 & F3Ia & 2009-05-06 & 1$\times$4$\times$491 & 664 & 12696 &  -2.98$\pm$2.45 & 3.121E-01 & ND \\
$B$~Vel & 74180 & F3Ia & 2010-03-08 & 3$\times$4$\times$80 & 1789 & 33932 & 1.47$\pm$1.94 & 9.557E-01 & ND \\
$\delta$~Cep & 213306 & F5Iab & 2009-10-02 & 3$\times$4$\times$97 & 1726 & 37934 & -0.25$\pm$0.58 &  9.357E-01 & ND\\
$\gamma$~Cyg & 194093 & F8Ib & 2009-09-06 & $3\times4\times$23 & 2738 & 75220 & -0.24$\pm$0.40 & 6.822E-01 & ND\\
$\delta$~CMa & 54605 & F8Iab & 2009-09-27 & 3$\times$4$\times$64 & 2754 & 74751 & 0.37$\pm$0.51 & 8.030E-01 & ND \\
83~Vir & 119605 & G0Ib-IIa & 2009-05-05 & 1$\times$4$\times$185 & 1448 & 42557 & -0.70$\pm$0.72 & 4.022E-01 & ND \\
$\zeta$~Gem & 52973 & G0Ibv & 2009-09-10 & 3$\times$4$\times$100 & 2346 & 54880 & 0.29$\pm$0.39 & 5.453E-01 & ND \\
$F$~Hya & 74395 & G1Ib & 2010-01-28 & 2$\times$4$\times$201 & 2553 & 75160 &  0.29$\pm$0.39 & 8.439E-01 & ND \\
$\beta$~Pyx & 74006 & G7Ib-II & 2009-11-24 & 3$\times$4$\times$88 & 2706 & 89445 & 0.93$\pm$0.39 & 5.832E-01 & ND \\
HR~3612 & 77912 & G8Iab & 2009-05-05 & 4$\times$4$\times$72 & 1878 & 60586 & -0.05$\pm$0.18 & 9.165E-01 & ND \\
HR~5009 & 115337& K0Ib & 2009-05-07 & 1$\times$4$\times$320 & 681 & 21839 & -2.73$\pm$1.34 &  1.091E-02 & ND \\
$\epsilon$~Peg & 206778 & K2Ib & 2008-12-20 & 5$\times$4$\times$15 & 479 & 30740 &  0.03$\pm$0.81 & 4.128E-01 & ND \\ 
$\pi$~Pup & 56855 & K3Ib & 2009-11-28 & 2$\times$4$\times$23 & 1641 & 52060 & -1.24$\pm$0.42 & 4.962E-01 & ND\\
$\alpha$~Sco & 148478 & M1.5Iab-b & 2010-01-25 & 2$\times$4$\times$4 & 2165 & 58501 & 0.69$\pm$0.37 & 4.634E-01 & ND\\
\hline
\hline
\end{tabular}\\
\noindent$^1$The data presented in this paper are the co-addition of observations taken on 2009-09-28, 2009-10-02, and 2009-10-07.
\label{obs_tab}
\end{table*}

\section{Magnetic field diagnosis}\label{mag_fields}
\begin{figure*}
\centering
\includegraphics[width=6.8in]{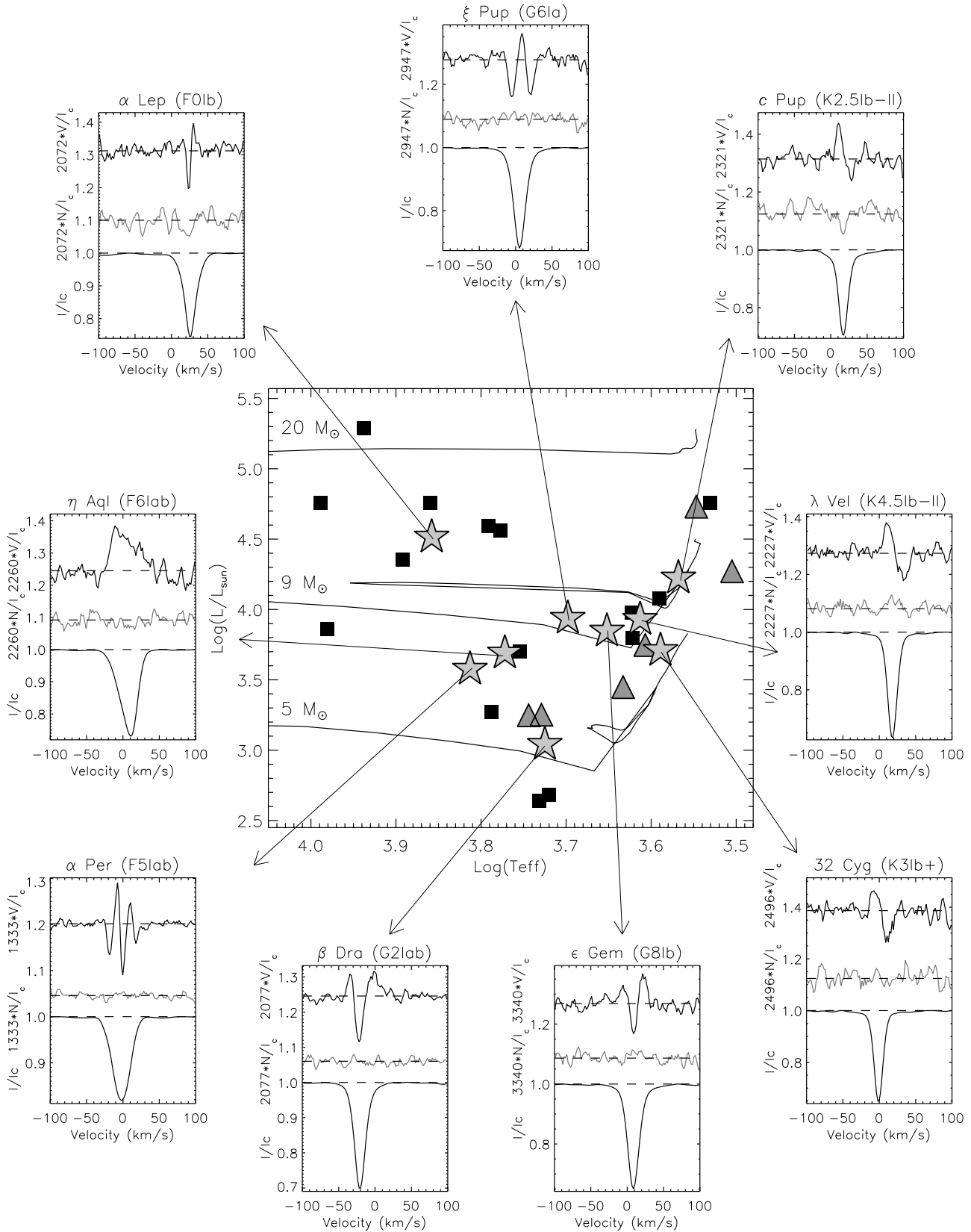}
\caption{Hertzsprung-Russell diagram (center frame) showing all supergiants observed as part of our survey for which parameters were available, with evolutionary tracks of \citet{schal92}. Black squares indicate stars for which no clear Zeeman signature is found, grey triangles indicate stars for which a Zeeman signature is suspected, and grey stars indicate stars with a clear Zeeman signature. Surrounding the HR diagram are illustrative LSD Stokes~$I$ (bottom), Stokes~$V$ (top) and diagnostic null (middle) profiles for the 9 stars with unambiguously detected Stokes~$V$ signatures.}
\label{hr_diag}
\end{figure*}

In order to increase the S/N and detect weak Zeeman signatures in the circularly polarized spectra, we applied the Least-Squares Deconvolution \citep[LSD;][]{don97} procedure to all polarimetric observations. The line masks used in this analysis were constructed for photospheric lines using atomic data from the Vienna Atomic Line Database \citep[VALD;][]{rya97, kupka99} with line depths computed assuming solar abundances and LTE. Only lines with predicted depths greater than 10 percent of the continuum flux were included in the line masks. This resulted in, for example, about 4000 lines included in the mask for a G2 star. The final result of the LSD analysis of a sequence of observations is a single, mean circular polarisation profile (LSD Stokes~$V$), a mean diagnostic null profile (LSD $N$), and a mean un-polarized profile (LSD Stokes~$I$). This resulted in an increase in the S/N by a factor of 21 to 68, roughly varying as the square root of the number of lines in the mask. All LSD profiles were produced on a spectral grid with a velocity bin of 1.8~km\,s$^{-1}$. 

\begin{figure*}
\centering
\includegraphics[width=6.8in]{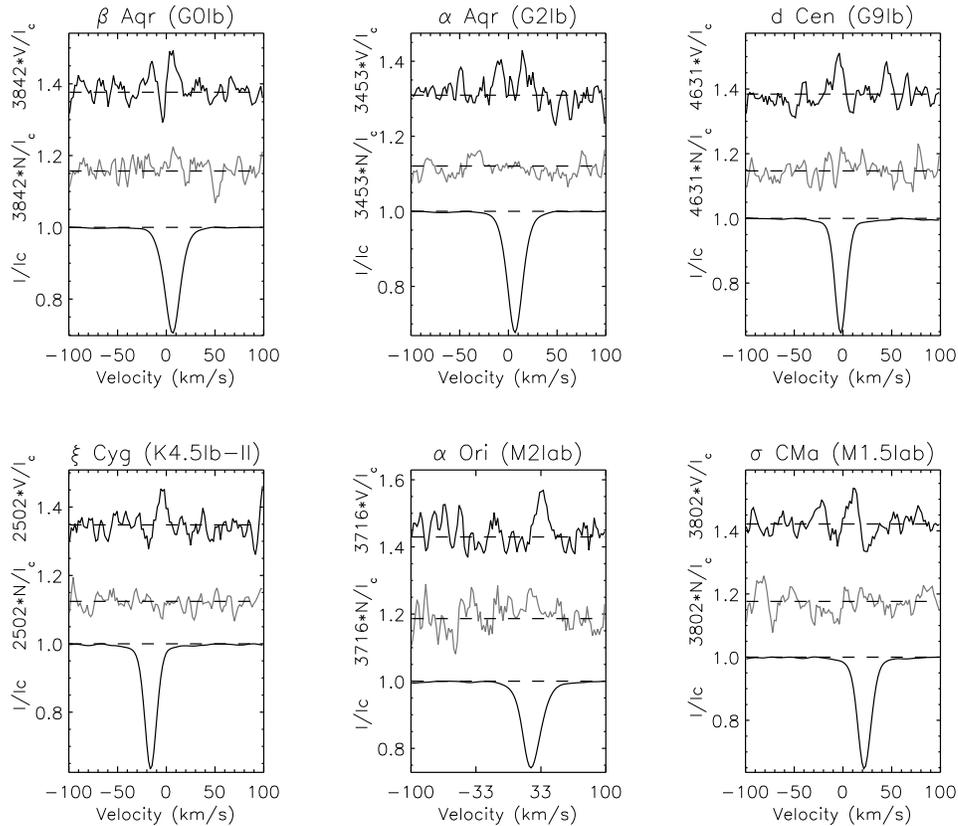}
\caption{LSD Stokes~$I$ (bottom), Stokes~$V$ (top) and diagnostic null (middle) profiles for stars with suspected Zeeman signatures detected in Stokes~$V$ (based on an excess of Stokes~$V$ signal at the position of the stellar mean spectral line), which still result in non-detections based on the detection criteria of \citep{don97}.}
\label{grid1}
\end{figure*}

Fig.~\ref{hr_diag} illustrates the LSD profiles of 9 stars of our sample for which magnetic fields were unambiguously detected, showing the often complex Zeeman signatures across the spectral lines of supergiants from early-F to mid-K spectral type and their placement on the HR diagram. Temperatures and luminosities were compiled from various published catalogues \citep{cayrel01, gray01, kov07, lyu10}, or otherwise derived using $UBV$ photometry from the Bright Star Catalogue \citep{hoff91} using transformations of \citet{flower96} and \citet{bessell98}. Six additional stars in which no statistically significant Zeeman detection is achieved but in which the presence of magnetic field is suspected (based on an excess of Stokes~$V$ signal at the position of the stellar mean spectral line) are shown in Fig.~\ref{grid1}. Finally, the remaining stars presenting no evidence of a Zeeman signature are illustrated in Fig.~\ref{grid2}.

The longitudinal field measurements reported in Table~\ref{obs_tab} were measured from the LSD Stokes~$I$ and $V$ profiles in the manner described by \citet{silv09}, with the integration carried out within the limits of the Stokes~$I$ line profile. We stress that while the longitudinal field provides a useful measure of the line-of-sight component of the field, we do not use it as the primary diagnostic of the presence of a magnetic field. This is because a large variety of magnetic configurations can produce a null longitudinal field (as is apparent from our measurements in Table~\ref{obs_tab}), while many of these configurations will generate a detectable Stokes~$V$ signature in the velocity-resolved line profiles that we observed, as is evident in Fig.~\ref{hr_diag}.

We find an increase in the number of detectable Stokes~$V$ signatures for observations of higher S/N (the peak S/N of the unpolarized spectra in the $V$-band ranges from 700 to 3300). We find only 4 detections ($\sim$40 percent of our detected number of stars) with peak S/N of 2500 or less (the bottom half of our sample), while we find 5 detections ($\sim$60 percent of our detected number of stars) with peak S/N in the top half of our sample. However, we do detect fields in stars with lower S/N, the lowest being $\alpha$~Per with a peak S/N of 1509 in the $V$-band. Nevertheless, based on our own experience and that of \cite{aur10} it appears that achieving a S/N of 2500 or higher is best suited to detecting and characterising magnetic fields in supergiants using the methods described here. 

\begin{figure*}
\centering
\includegraphics[width=6.8in]{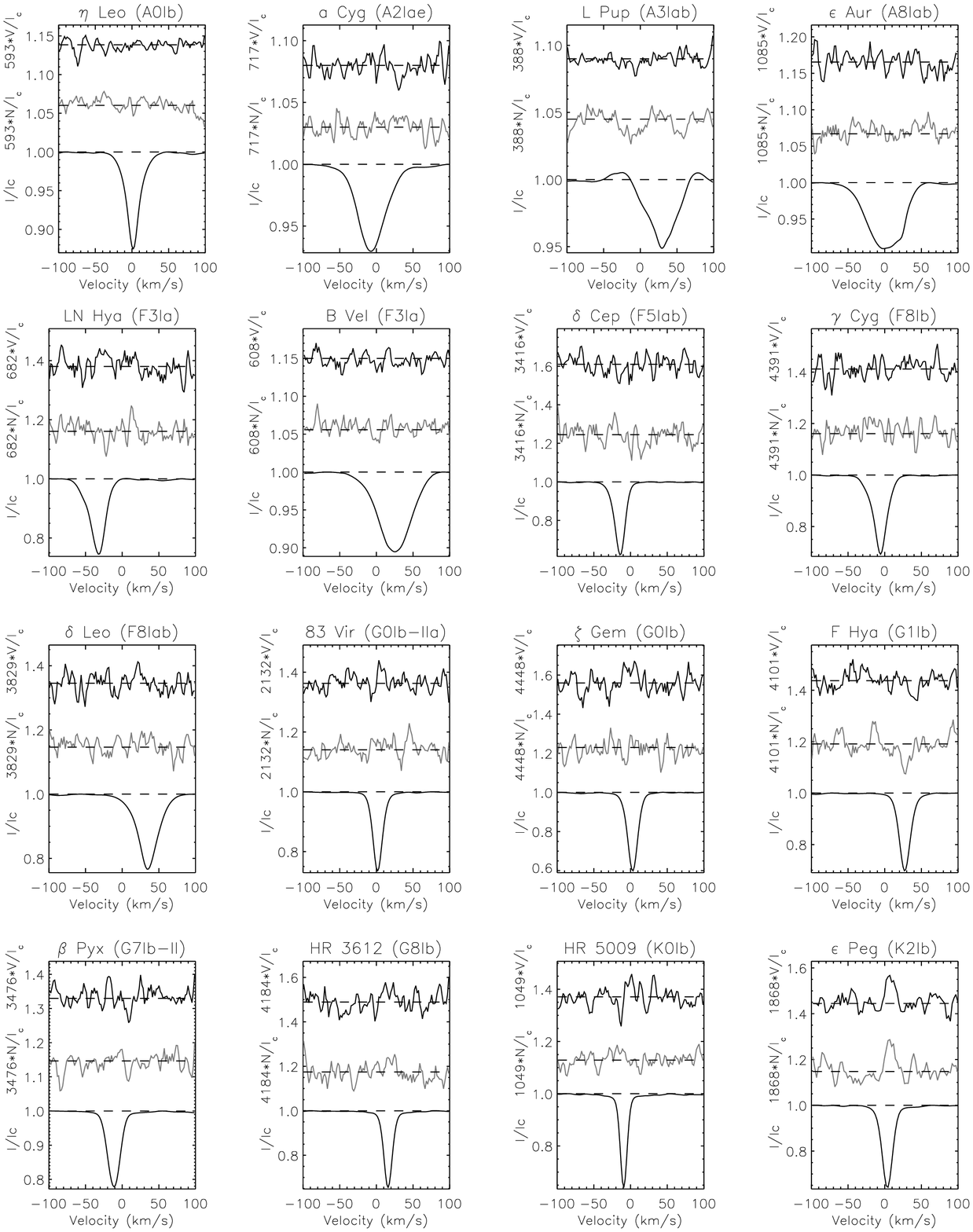}
\caption{LSD Stokes~$I$ (bottom), Stokes~$V$ (top) and diagnostic null (middle) profiles for stars with no evidence of a Zeeman signature in Stokes~$V$.}
\label{grid2}
\end{figure*}

\section{Results}\label{disc_sec}
Our investigation of late-type supergiants shows that many host detectable Stokes~$V$ Zeeman signatures. The signatures are frequently complex, and the associated longitudinal magnetic field are generally weaker than 1~G. We have detected usually unambiguous Zeeman signatures in 9 stars of our sample, with 6 additional stars suspected to show signatures. The detected stars span a large range of physical characteristics, with the most massive star detected being $\alpha$~Lep (F0Ib, $\sim$15\,$M_{\odot}$). This star also represents the hottest star in our sample with a detectable circular polarisation Zeeman signature ($T_{\rm eff}\sim7200$\,K). The lowest mass star in our sample with a detectable Zeeman signature is $\beta$~Dra (G2Iab, $\sim$5\,$M_{\odot}$). Based on our adopted effective temperatures, the coolest star with a detection is $c$~Pup ($T_{\rm eff}\sim3700$\,K), but we also have obtained detections in other cool K-type stars such as $\lambda$~Vel.

As is evident from the HR diagram in Fig.~\ref{hr_diag}, we obtain a high incidence fraction for stars with spectral types F ($\sim$40 percent; total of 8 stars observed), G ($\sim$30 percent; total of 11 stars observed), and K ($\sim$40 percent; total of 7 stars observed). For M-type supergiants this fraction is more uncertain (with a total of just 3 stars observed); with two suspected detections the incidence fraction may be as high 67 percent. We currently have not obtained any detections in hotter A-type stars (total of 4 stars observed).

Prior to November 2009, ESPaDOnS was known to suffer from (variable) polarisation cross-talk at a level of 1-4 percent. During the period when most of the observations described here were acquired, the primary source of this cross-talk was the ESPaDOnS atmospheric dispersion corrector \citep[ADC;][]{barr10}. In November 2009, a new ADC was installed in front of the ESPaDOnS polarimetric module, and the measured cross-talk since that time has remained stable at a level of about 0.6 percent. The repeatability of our Zeeman detections (both with ESPaDOnS, before and after resolution of the cross-talk issue, as well as with Narval), the existence of both detections and non-detections within our sample, the diversity and complexity of the detected signatures, and the lack of any published evidence for strong, coherent linear polarisation within the metallic absorption lines of late-type supergiants, makes us confident that the Stokes~$V$ signatures detected within this survey are not significantly affected by cross-talk.

\begin{figure}
\centering
\includegraphics[width=2.5in]{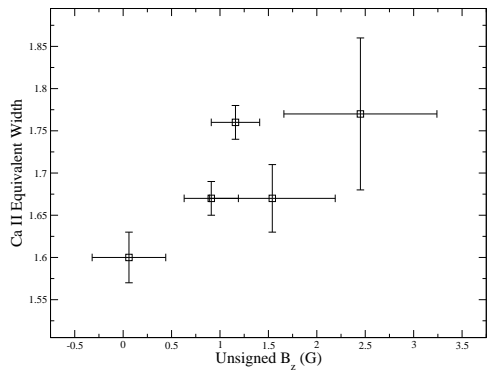}
\caption{Ca~{\sc ii} K line core equivalent width measurements for $\beta$~Dra as a function of the unsigned longitudinal magnetic field (B$_{\rm z}$).}
\label{ca2_var}
\end{figure}

Our current dataset reveals no clear differences between the classical optical activity indicators (such as Ca~{\sc ii} H\&K emission or H$\alpha$ emission) of those stars for which we detect fields and those for which we do not. Additionally, we find no obvious differences between the rotational velocities of the stars with magnetic detections and those without (using $v\sin i$ measurements of \citet{demed02}). However, there does appear to be a weak correlation between the unsigned longitudinal field strength and Ca~{\sc ii} core equivalent width measurements for some stars in which fields are detected on multiple dates, such as $\beta$~Dra and $\epsilon$~Gem (this correlation for $\beta$~Dra is shown in Fig.~\ref{ca2_var}).

Historically, a number of searches for magnetic fields in late-type supergiants have been published \citep[e.g.][]{bor81, scholz81, plach05}. Essentially, our direct detections of magnetic fields in supergiants - in particular our typical observation of sub-1 G longitudinal fields - are not consistent with any of the reports claimed in those papers.

\begin{figure}
\centering
\includegraphics[width=3.2in]{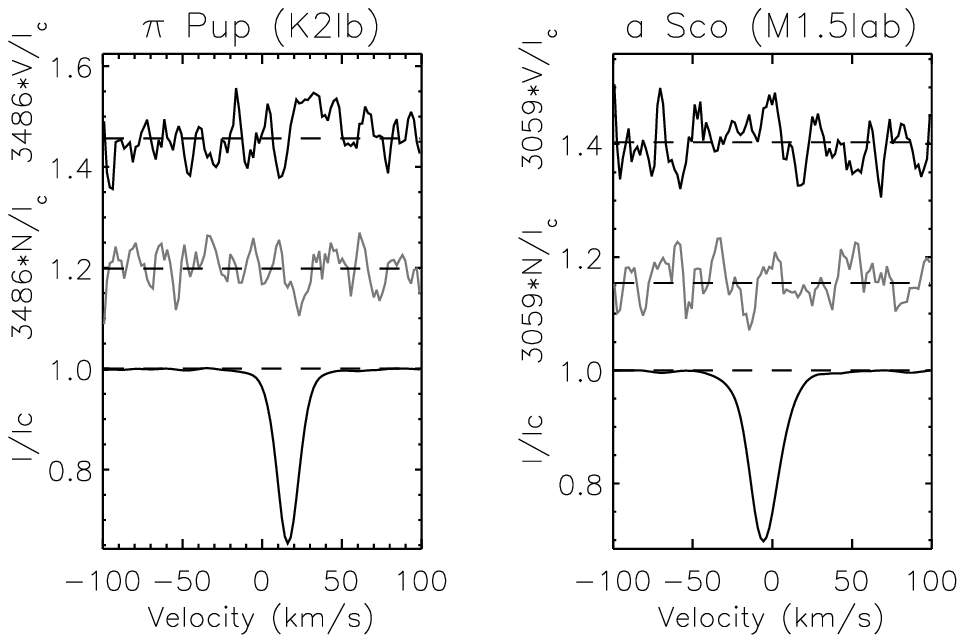}
\contcaption{LSD Stokes~$I$ (bottom), Stokes~$V$ (top) and diagnostic null (middle) profiles for stars with no evidence of a Zeeman signature in Stokes~$V$.}
\end{figure} 

\begin{figure}
\centering
\includegraphics[width=3.2in]{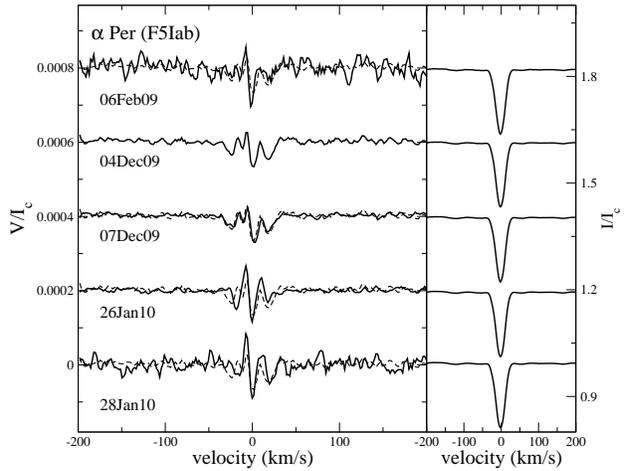}
\caption{Stokes~$V$ profiles (left, solid lines) and Stokes~$I$ (right) of $\alpha$~Per for the nights indicated. Profiles are vertically offset for display purposes. The dashed line corresponds to the observation obtained on 04 Dec. 2009, shifted to the position of each night in order to highlight changes in the profile.}
\label{hd20902_var}
\end{figure}

A few of the stars in our sample were previously investigated for magnetic fields. Several previous magnetic analyses of $\alpha$~Per were unable to detect a significant field, with the lowest uncertainty that of \citet{shor02} with $\langle B_{\rm z}\rangle=1\pm2$\,G and another measurement by \citet{bor73} ($B_z=0\pm49$\,G). Our 5 observations of this star span almost a full year, show some variability (see Fig.~\ref{hd20902_var}) and confirm the presence of a weak longitudinal field ($\sim 1$~G). While this field is weak, this star exhibits the strongest and most complex Stokes~$V$ profile of all observed stars, with an amplitude larger by a factor of $\sim$2.

A magnetic analysis $\eta$~Aql was carried out by \citet{wade02} to investigate the reported detections by \citet{plach00}. Wade et al. found an average unsigned longitudinal field strength of $10$\,G or less and a typical uncertainty of 5\,G, while Plachinda claimed maximum field strengths of $\sim$100\,G, varying with the pulsation period. Our observation of this star does show a clear signature (see Fig.~\ref{hr_diag}), corresponding to a null longitudinal field measurement ($B_z=-0.23\pm0.75$\,G) consistent with the low values of Wade et al. near the same pulsation phase (phase 0.76). We note the possibility that the velocity field due to pulsation is responsible for the unusual shape of the observed Stokes~$V$ profile.

\begin{figure}
\centering
\includegraphics[width=3.2in]{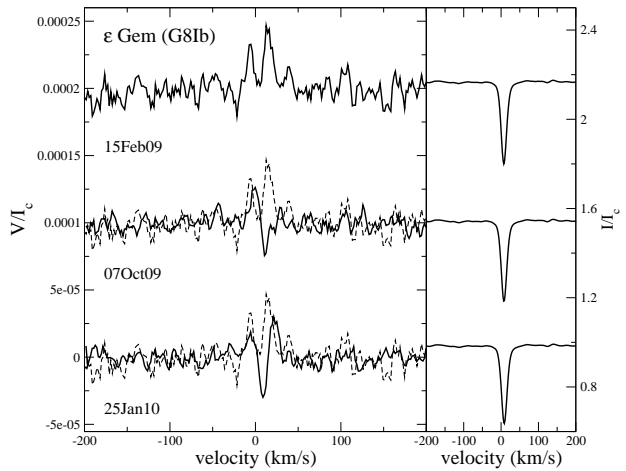}
\caption{Stokes~$V$ profiles (left, solid lines) and Stokes~$I$ (right) of $\epsilon$~Gem for the nights indicated. Profiles are vertically offset for display purposes. The dashed line corresponds to the observation obtained on 15 Feb. 2009, shifted to the position of each night in order to highlight changes in the profile.}
\label{hd48329_var}
\end{figure}

$\epsilon$~Gem was investigated by \citet{plach05}, who reported the detection of a variable field with unsigned field strengths ranging from $\sim$2 to $\sim$50\,G, with a typical uncertainty $\sim$5\,G. While we do find a clear Zeeman signature in Stokes~$V$, our measured longitudinal fields of this star are significantly lower than the detections claimed by \citet{plach05}. The observed Stokes~$V$ profile of $\epsilon$~Gem appears to vary on the order of months, as illustrated in Fig.~\ref{hd48329_var}.

Several supergiants for which we have detected a Zeeman signature are also known pulsators or variable stars (such as $\alpha$~Lep \citep{ark05}, $\eta$~Aql \citep{kiss00}, $\xi$~Pup \citep{koen02}). Curiously, $\alpha$~Per is a non-pulsating star that clearly lies within the instability strip \citep{kell08} with evidence of circumstellar shells \citep[e.g.][]{neil08} and a stronger Zeeman signature.

A few of the stars in our sample fall into the ``hybrid" chromosphere category, such as $\alpha$~Aqr, $\beta$~Aqr, and $c$~Pup. All of these stars have weak but detectable X-ray emission \citep{ayres05a}. While neither $\alpha$~Aqr or $\beta$~Aqr have detected Zeeman signatures in Stokes~$V$, our observation do show complex Stokes~$V$ profiles that are suggestive of a magnetic field. Comparing with the active stars $\beta$~Dra or 32~Cyg (for which UV prominences were detected by \citet{schro83}), we see no clear differences in the strengths of their Stokes~$V$ signatures that would reflect differences in their magnetic fields. We also observe a clear signature in $\lambda$~Vel, a ``non-coronal" star with a typical cool chromosphere \citep{carp98}.

\section{Conclusions}
Our survey of magnetic fields in massive, late-type supergiants reveals detectable Zeeman Stokes~$V$ signatures in the LSD profiles of about one-third of our sample of more than 30 stars. The signatures are sometimes complex and correspond to longitudinal magnetic fields generally below 1\,G. These characteristics suggest topologically complex magnetic fields, presumably generated by dynamo action. Given the high rate of incidence, it may well be that magnetic fields are excited in all cool supergiants. Nevertheless, our failure to detect fields in some stars for which very high S/N observations were acquired points to a large range of field strengths or complexities. In fact, \citet{aur10} report the detection of a clear Zeeman signature in Stokes~$V$ spectra of the cool M2 supergiant Betelgeuse, with a S/N almost twice has high as typically achieved in this study. This leads us to believe that a S/N approximately twice as high as that achieved here would be valuable for a more complete assessment of field incidence. An additional interesting result of this study is the detection of Zeeman signatures in hybrid and non-coronal supergiants - signatures with strength and structure similar to those observed in active supergiants. As similar magnetic fields are therefore inferred to exist in all three classes of stars, this suggests that the magnetic properties of these different classes are similar and that other phenomena (e.g. attenuation by cool winds or extended chromospheres) are likely the cause of the observed differences in their activity. 

\section*{ACKNOWLEDGMENTS}
The authors thank Dr. John Landstreet for helpful discussion and effective motivation, and the referee Dr. Tony Moffat for constructive advice. JHG acknowledges financial support in the form of an Ontario Graduate Scholarship. GAW and DAH acknowledge Discovery Grant support from the Natural Science and Engineering Research Council of Canada. This research has made use of the SIMBAD database, operated at CDS, Strasbourg, France.

\label{lastpage}
\end{document}